# Logistic Models of Fractal Dimension Growth for Spatio-temporal Dynamics of Urban Morphology

Yanguang Chen

(Department of Geography, College of Environmental Sciences, Peking University, Beijing 100871, PRC. Email: chenyg@pku.edu.cn)

**Abstract**: Urban form and growth can be described with fractal dimension, which is a measurement of space filling of urban evolution. Based on empirical analyses, a discovery is made that the time series of fractal dimension of urban form can be treated as a sigmoid function of time. Among various sigmoid functions, the logistic function is the most probable selection. However, how to use the model of fractal dimension growth to explain and predict urban growth is a pending problem remaining to be solved. This paper is devoted to modeling fractal dimension evolution of different types of cities. A interesting discovery is as follows: for the cities in developed countries such as UK, USA and Israel, the comparable fractal dimension values of a city's morphology in different years can be fitted to the logistic function; while for the cities in developing countries such as China, the fractal dimension data of urban form can be fitted to a quadratic logistic function. A generalized logistic function is thus proposed to model fractal dimension growth of urban form. The logistic model can also be employed to characterize the change of the dimensions of multifractals. The generalized logistic models can be used to predict the missing values of fractal dimension, to estimate the growth rate of the dimension, and thus to reveal the spatio-temporal process and pattern of a city's growth. Especially, the models lay a foundation for researching the correlation between urban form and urbanization and for developing the theory of spatial replacement dynamics.

**Key words**: fractal; fractal dimension; logistic function; urban form and growth; replacement dynamics; Beijing

# 1. Introduction

A scientific research should proceed first by describing how things work and later by understanding why (Gordon, 2005; Henry, 2002). The describing is mainly by mathematics and measurement, and the understanding is mainly by observation, experience, and experiments (Henry, 2002). For a geographical phenomenon with characteristic scales, we can describe it using the conventional measures such as length, area, volume, and density. In this case, the traditional mathematical methods can be employed to make quantitative analyses by describing a set of measurements. However, urban form is a kind of scale-free phenomenon, which shows no characteristic scale, and cannot be described by conventional measurement and traditional mathematics. Fractal geometry is an efficient tool of scaling analysis and can be used to describe scale-free geographical distributions and evolution. In fact, fractal dimension proved to be more efficient to describe urban form than the common measures such as length, area, and density (Batty and Longley, 1994; Frankhauser, 1998). Nowadays, fractal dimension has been used as varied form indices of cities (Batty and Kim, 1992; Chen, 2011; De Keersmaecker et al, 2003; Feng and Chen, 2010; Frankhauser, 2008; Thomas *et al*, 2007; Thomas *et al*, 2008; Thomas *et al*, 2010), or employed to research the spatial dynamics of urban form (Benguigui *et al*, 2001a; Benguigui *et al*, 2006). A single fractal dimension value says a little thing. The function of a measurement or parameter rests with comparison of values, including year-to-year comparison and one-to-one comparison. By comparison of fractal dimension values of different cities in given time or a given city in different years, we can reveal the geographical spatio-temporal information about urban evolution from a city as a system or a system of cities.

Urban evolution indicates a complex space-filling process and pattern, and urban growth can be associated with spatial dynamics of urban morphology. In order to research a growing process of a fractal city, we should compute its fractal dimension year by year. Bring the dimension values of different years into comparison, we can find some regularity of urban growth. However, scattered fractal dimension values cannot reflect the evolution regularity of cities. In order to reveal the hidden law of urban evolution, we should describe a set of fractal dimension values. Mathematical modeling of fractal dimension growth will help us find the deep structure of urban form. In practice, using the mathematical model, we can estimate the missing value of fractal dimension in the past (due to absence of observed data), predict the maximum fractal dimension indicative of the capacity of space filling in the future, or even judge when the rate of fractal dimension growth reached or will reach its peak value. The fractal dimension increase suggests urban growth, so the model can be used to make dynamical analysis on cities. Of course, description differs from understanding. In order to explain urban spatial dynamics, we must rely heavily on observation, experience, computer simulation, or numerical computation.

A process of urban growth is associated with urbanization, which comprises urban form, urban ecology, urbanism, and urban system (Knox and Marston, 2009). By analogy with the model of urbanization level and a number of empirical analyses, I find that fractal dimension growth takes on sigmoid curves and can be modeled with the logistic function or some similar function (Chen, 2012). The logistic model of fractal dimension can be utilized to predict urban growth. In theory, the logistic map can be employed to explore the complex spatial dynamics of city development or urban evolution. However, how to make empirical analysis using this kind of function is a pending problem remaining to be solved. The rest parts of the paper will be organized as follows. First, I present the logistic model of fractal dimension by qualitative analysis, and propose two approaches of quantitative analyses to estimating the parameter values. Second, one dataset of London from Batty and Longley (1994), three datasets of Tel Aviv from Benguigui *et al* (2000), and one dataset of Baltimore from Shen (2002) are employed to testify the model. Third, several questions will be discussed, and a quadratic logistic model will be proposed for the cities of developing countries. Finally, the article is concluded by summarizing the main points of this study.

## 2. Fractal dimension and sigmoid function

### 2.1 Basic postulates and model

If some kind of measure of a system has clear upper limit and lower limit, and if growth of the system is not of uniform speed, the growing course of the measure always takes on the S-shaped curve. Formally, the curve can be abstracted as what is called *sigmoid function*. Sometimes, the sigmoid function is called *squashing function* (Chen, 2012; Gershenfeld, 1999). Pressed by the upper limit and withstood by the lower limit, a line will be deformed and twisted into S shape.

Thus a straight line will change to a sigmoid curve. The family of sigmoid functions includes various functions such as the ordinary arc-tangent, the hyperbolic tangent, and the error function, the Gompertz function, and the generalized logistic function (say, Boltzmann's equation, quadratic logistic function). Among varied sigmoid functions, the logistic function is very familiar to scientists and engineers. So, sometimes, the sigmoid function is equated with the logistic function (Mitchell, 1997).

As we know, the level of urbanization ranging from 0 (or 0%) to 1 (or 100%) can often be modeled by the general logistic function since it has clear upper and lower limits (Chen, 2009; United Nation, 2004). In fact, in urban studies, fractal dimension of urban form and growth based on 2-dimensional digital maps always comes between 0 and 2 (Batty and Longley, 1994; Chen, 2010). In other words, the fractal dimension has clear upper limit ($D_{max}$=2) and lower limit ($D_{min}$=0). So, it is possible that fractal dimension of urban form follows the law of logistic growth, and can be described with the typical sigmoid function. The model can be expressed as

$$D(t) = \frac{D_{max}}{1+(D_{max}/D_0 - 1)e^{-kt}} \quad \text{or} \quad D(n) = \frac{D_{max}}{1+Ae^{-k(n-n_0)}}, \quad (1)$$

where $t$ refers to time order (0, 1, 2,…), and $n$ to year, $n_0$ to the initial year, $D(t)$ or $D(n)$ denotes the fractal dimension in the $t$th time or the year of $n$, $D_0$ is the fractal dimension in the initial time/year, $D_{max}$≤2 indicates the maximum of the fractal dimension (the terminal or capacity of fractal parameter), $A$ refers to a parameter, and $k$ to the original growth rate of fractal dimension. The parameter and variable relationships are as follows

$$D(t) = D(n), \quad A = \frac{D_{max}}{D_0} - 1, \quad t = n - n_0.$$

In fact, it has been demonstrated that equation (1) can be derived from a system of nonlinear equations on urban spatial dynamics (Chen, 2012).

So far, we have had three major evidences for the choice of the model. The first is theoretical evidence, that is, equation (1) can be derived from urban spatial replacement equations (Chen, 2012). The second is mathematical evidence, that is, the model can be directed by the squashing principle of developing space (Gershenfeld, 1999). The third is the circumstantial evidence, that is, it can be proposed by analogy with the model of urbanization curve (United Nation, 2004). In next section, I will present the fourth evidence, that is, the empirical results that supported the mathematical modeling.

**2.2 Two methods of parameter estimation**

The parameters of a linear mathematical model can be estimated by the ordinary least squares (OLS) method. However, it is not convenient to estimate the parameters of nonlinear model by OLS. If we use the method of nonlinear fit (NLF) based on Jacobian iterations by means of Matlab or Mathcad, the estimated parameter values are not so stable. If a nonlinear equation has one or two parameters, it can be easily linearized for OLS computation by some kind of mathematical method, especially logarithmic transform. However, if a nonlinear model has three or more parameters, generally speaking, it is difficult to transform it into a linear form. Unfortunately, equation (1) is a three-parameter nonlinear function. Since OLS is very simple and familiar to geographers, I recommend two methods based on OLS for estimating the parameters of the logistic model.

The first method can be termed "*bivariate nonlinear auto-regression* (BNAR)" based on OLS. Taking derivative of equation (1) yields

$$\frac{dD(t)}{dt} = kD(t)[1 - \frac{D(t)}{D_{max}}], \qquad (2)$$

which can be discretized as

$$\frac{\Delta D_t}{\Delta t} \cong kD_{t-1}(1 - \frac{D_{t-1}}{D_{max}}). \qquad (3)$$

where $D_t$ is the discrete form of $D(t)$, and "$\cong$" suggests some error resulting from the conversion from continuous to discrete equation. Thus we have a bivariable nonlinear auto-regression model

$$D_t = (1 + k\Delta t)D_{t-1} - \frac{k\Delta t}{D_{max}}D_{t-1}^2 = \mu D_{t-1} - \eta D_{t-1}^2. \qquad (4)$$

Taking $D_{t-1}$ and $D_{t-1}^2$ as two independent variables, and $D_t$ as a dependent variable, we can implement a multivariate linear regression analysis by letting the constant item equal zero. A least square calculation will give

$$\mu = 1 + k\Delta t, \quad \eta = \frac{k\Delta t}{D_{max}}.$$

This implies

$$D_{max} = \frac{\mu - 1}{\eta}. \tag{5}$$

Substituting equation (5) into equation (1), we can make a logarithmic transform and the result is

$$\ln(\frac{D_{max}}{D(t)} - 1) = \ln(\frac{D_{max}}{D_0} - 1) - kt = \ln A - kt. \tag{6}$$

This is just a log-linear equation. By linear regression analysis, we can estimate $A$ and $k$ values.

The second method can be named "*goodness-of-fit search* (GOFS)". The goodness of fit equals the correlation coefficient square ($R^2$) for the ordinary linear regression. If we don't know the $D_{max}$ value, we can estimate an initial value for it, say, let $D_{max}(0)=2$. Then, we can use equation (6) to make a least square calculation which give a $R^2$ value, and the result can be notated as $R^2(0)$. Then, change $D_{max}(0)$ value, say, decrease it to 1.9, notated as $D_{max}(1)=1.9$. In this case, we have another $R^2$ value, notated as $R^2(1)$. If $R^2(1) > R^2(0)$, decrease $D_{max}$ again (take $D_{max}(2)<1.9$); if $R^2(1)<R^2(0)$, increase $D_{max}$ value (take $D_{max}(2)>1.9$). Thus we have $R^2(2)$. Continue this process until $R^2(i+1) \approx R^2(i)$ (step number $i=0, 1, 2,...$). You can make a Matlab programming and implement the GOFS process in Matlab's data environment. If you have no idea of Matlab programming, you can make use of MS Excel. Input the data into an Excel sheet, insert a scatter-plot, add a trend line onto the plot and have the $R^2$ value as well as the formula shown. Then you can perform the computation of GOFS step by step by using Excel's tools for data analysis.

The BNAR approach requires the sample path of time series of fractal dimension to be long enough and the temporal variable to be continuous or the time intervals of sampling are equal. In practice, however, this requirement cannot always be met. The GOFS approach has no special precondition but a set of data. Thus its sphere of application is larger than that of BNAR. GOFS seem to be likened to a "brute force" algorithm, but it is an efficient method, and the results are always satisfying. If you are familiar with MS Excel enough, it will take you no more than 1 minute to evaluate the three parameters approximately.

Scientific research is always involved with three worlds: *real world*, *mathematical world*, and *computational world* (Casti, 1996). In theory, the mathematical world can be directly associated with the real world; however, in empirical studies, the mathematical world is connected to the real world by way of the computational world. Both the real and mathematical worlds are objective, but the there is subjectivity in the computational world. Cities are defined in the real world, logistic reasoning and derivation are defined in the mathematical world, and algorithms, data processing, statistical analysis, and all that, are defined in the computational world. We can develop a fractal city model in the mathematical world based on certain postulates. But if we try to apply the mathematical model to a real city, we must utilize the computational world because the work of parameter estimation is fulfilled by certain algorithm and data processing method. For example, if we use the box-counting method to make measurement, the largest box may be fixed or unfixed; if we estimate the box dimension, we can use the least squares method (LSM), or the maximum likelihood method (MLM), or the major axis method (MAM). Different sampling methods and algorithms will results in different values of fractal dimension. For a mathematical modeling result of a natural or social phenomenon, in a sense, we can say it is good or bad, but cannot say it is right or wrong (Gabaix and Ioannides, 2004). Next, I will present three examples to illustrate the logistic models of fractal dimension growth of cities using the data from literature.

## 3. Empirical evidences

### 3.1 Case 1: London

The first case is on the capital and the largest city of the United Kingdom (UK), and the fractal dimension dataset of urban form comes from Batty and Longley (1994), who processed the data from Abercrombie (1945) and Doxiadis (1968) and gave the fractal dimension values from 1820 to 1962 (Table 1). The fractal parameter is the *radial dimension* based on area-radius scaling method (Frankhauser and Sadler, 1991). The fractal dimension in 1981 was estimated by Frankhauser (1994) (see also Batty and Longley, 1994). Because of the inconsistency of data "caliber", I show it on the plot as an "outlier" for reference only. The growth trend of the fractal dimension values based on Abercrombie (1945)/Doxiadis (1968) can be well modeled by the logistic function (Figure 1). Since the dataset is not large enough, and especially, the time intervals of sampling are not equal to one another, we should use the second method, GOFS, to estimate the parameters of the logistic model. A least square calculation of utilizing the data of Table 3 yields

$$\hat{D}(t) = \frac{1.812}{1 + 0.332 e^{-0.024t}},$$

where $t=n-1820$. The goodness of fit is about $R^2=0.857$. The estimation of parameters and the effect of data fit to model are fair and reasonable.

**Table 1. The fractal dimensions growth of London metropolis, 1820-1962**

| Year | Dimension | Year | Dimension | Year | Dimension | Year | Dimension |
|---|---|---|---|---|---|---|---|
| 1820 | 1.322 | 1860 | 1.415 | 1900 | 1.737 | 1939 | 1.791 |
| 1840 | 1.585 | 1880 | 1.700 | 1914 | 1.765 | 1962 | 1.774 |

**Source**: Batty and Longley, 1994, page 242.

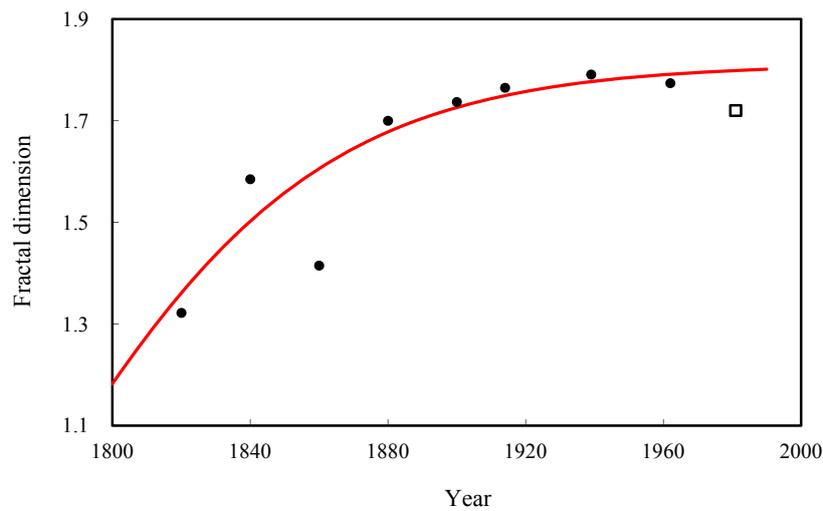

**Figure 1 The logistic patterns of fractal dimension growth of London, 1820-1962** (The data point for 1981 represented by a small square came from Frankhauser, 1994; the other data represented by round dots came from Batty and Longley, 1994)

### 3.2 Case 2: Tel Aviv

The second case is on the urban form of Tel Aviv, the second most populous city in Israel, and the datasets of fractal dimension comes from Benguigui et al (2000). The fractal parameter is the box dimension based on box-counting method. In order to determine when and where a city fractal is, Benguigui et al (2000) defined three study regions on the digital maps of Tel Aviv metropolis and estimated the fractal dimension in different years (from 1931 to 1991). Region 3 includes region 2, and region 2 contains region 1. The results of fractal dimension and the related parameters are listed in Table 2. By means of OLS, we can fit the logistic model, equation (1) to the three datasets. The calculations are listed in Table 3 and the effect of data points matching trend lines are displayed in Figure 2. The logistic model for region 1 is

$$\hat{D}(t) = \frac{2.000}{1+0.300e^{-0.018t}},$$

The correlation coefficient square is about $R^2=0.968$. The model for region 2 is

$$\hat{D}(t) = \frac{1.926}{1+0.387e^{-0.022t}},$$

The goodness of fit is about $R^2=0.991$. The model for region 3 is

$$\hat{D}(t) = \frac{1.823}{1+0.396e^{-0.025t}},$$

The coefficient of determination is about $R^2=0.989$. Using these models, we can estimate the years when the top speed of fractal dimension growth appeared and the peak value of dimension growth rate (Table 3).

**Table 2. The fractal dimensions growth of the three study regions in Tel Aviv, 1935-1991**

| Year | $D$ for Region1 | Std error | $D$ for Region2 | Std error | $D$ for Region3 | Std error |
|---|---|---|---|---|---|---|
| 1935 | 1.533 | 0.021 | 1.387 | 0.031 | 1.312 | 0.041 |
| 1942 | 1.607 | 0.018 | 1.434 | 0.027 | 1.341 | 0.044 |
| 1952 | 1.644 | 0.017 | 1.546 | 0.038 | 1.477 | 0.050 |
| 1962 | 1.672 | 0.025 | --- | --- | --- | --- |
| 1971 | 1.695 | 0.025 | 1.621 | 0.038 | 1.565 | 0.052 |
| 1978 | 1.773 | 0.014 | 1.679 | 0.026 | 1.600 | 0.041 |
| 1985 | 1.787 | 0.014 | 1.707 | 0.026 | 1.632 | 0.037 |
| 1991 | 1.809 | 0.014 | 1.733 | 0.025 | 1.667 | 0.037 |

**Source**: Benguigui *et al*, 2000.

**Table 3. The estimated values of the parameters in the logistical models for Tel Aviv**

| Parameter/Remark | Parameter value | | |
|---|---|---|---|
| | Region 1 | Region 2 | Region 3 |
| Capacity $D_{max}$ | 2.000 | 1.926 | 1.823 |
| Coefficient $A$ | 0.300 | 0.387 | 0.396 |
| Original growth rate $k$ | 0.018 | 0.0221 | 0.0250 |
| Goodness of fit $R^2$ | 0.962 | 0.991 | 0.989 |
| Years of top speed for dimension | 1868-69 | 1891-92 | 1897-98 |
| Peak value of dimension growth rate | 0.0090 | 0.0106 | 0.0114 |
| Remarks on geographical scope | Central part | Central and northeast parts | Entire metropolis |

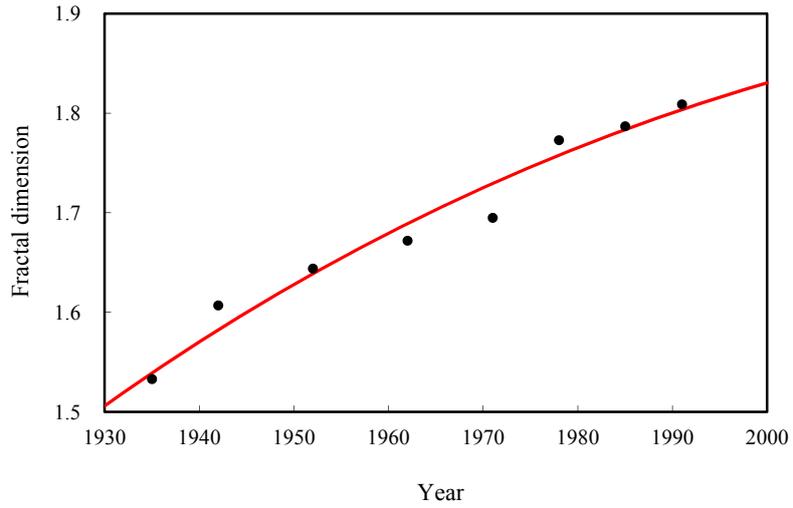

a. Regional 1

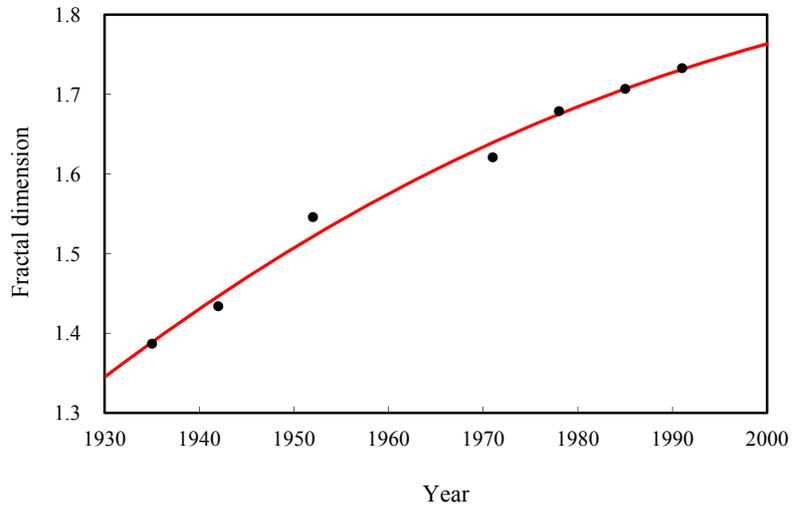

b. Regional 2

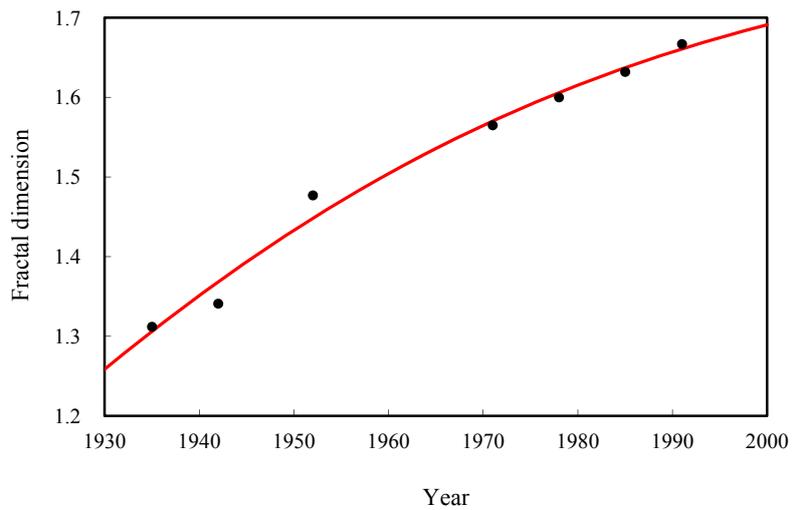

c. Regional 3

**Figure 2 The logistic patterns of fractal dimension growth of Tel Aviv, 1935-1991** (The solid dots

represent the empirical values of fractal dimension, and the line denotes the predicted values by the

logistic models)

The results can stimulate new thinking on where and when a city fractal is. For region l, if we seek the maximum value of the goodness of fit, the capacity parameter is about $D_{max}$=2.446>2. This suggests that the maximum of fractal dimension will exceed the upper limit of fractal dimension defined by the space in which a fractal exists. This is of anomaly because the box dimension cannot be greater than the Euclidean dimension of its embedding space ($d$=2). For guaranteeing the validity of fractal dimension, the capacity parameter is taken as $D_{max}$=2. For regions 2 and 3, the estimated values of the capacity parameter are less than the Euclidean dimension of the embedding space and are acceptable. What is more, the goodness of fit for region 2 is greater than those for regions 1 and 3. This seems to suggest that region 1 is too small and region 3 is too large for proper fractal dimension estimation. Relatively, the best size of a study area is region 2. If the study area is too large, the standard error of the fractal dimension will be too high ($\delta$>0.04); otherwise, the capacity parameter of the logistic model of fractal dimension will be not acceptable in logic. However, this does not mean that regions 1 and 3 are not useful for fractal study of Tel Aviv city. The question is very interesting, but it goes beyond the topic of this article and remains to be discussed in future.

### 3.3 Case 3: Baltimore

The third case is Baltimore, a city of northern Maryland on an arm of Chesapeake Bay, the United States of America (USA). The dataset of fractal dimension comes from Shen (2002), who estimated the fractal dimension of Baltimore's urban form from 1792 to 1992 (Table 4). Both Benguigui *et al* (2000) and Shen (2002) employed the box-counting method to estimate fractal dimension. However, there is a subtle distinction of technique. Benguigui *et al* (2000) defined a variable study area for Tel Aviv, while Shen (2002) defined a fixed study area for Baltimore. For fractal dimension estimation, it is easier to define a fixed study area than to define an objective variable area. The advantage of the fixed study area is that the fractal dimension values are more compatible for year-to-year comparison. However, nothing is perfect. For a fixed study area, the fractal dimension values of initial stages can be underestimated, while the dimension values of terminal stages can be overestimated (Chen, 2012). Maybe because of this, the capacity of fractal dimension of Baltimore cannot converge to a value less than 2. If we seek the maximum goodness of fit, the capacity parameter is about $D_{max}$=2.12. This value cannot be accepted because the box dimension cannot go beyond the Euclidean dimension of the embedding space, $d$=2; therefore, the maximum dimension is taken as $D_{max}$=2. Using OLS method, we can make a model as below:

$$\hat{D}(t) = \frac{2}{1+1.7365e^{-0.0118t}},$$

The determination coefficient is about $R^2$=0.9658 (Figure 3).

By means of this model, we judged that the maximum speed of fractal dimension growth appeared in 1839 or 1840, when the city size is about 100 thousands. This differs from the speed of population growth. According to the logistic modeling of urban population data from Shen (2002), the top speed of population growth appeared in 1900 or 1901. The time lag between the two top speeds is about 60 years. This suggests that when the population size exceeded 100 thousands, more high-rise buildings appeared in Baltimore so that the urban land use became more intensive than ever.

Table 4. The fractal dimensions growth of Baltimore, 1792-1992

| Year | Dimension | Year | Dimension | Year | Dimension | Year | Dimension |
| --- | --- | --- | --- | --- | --- | --- | --- |
| 1792 | 0.6641 | 1878 | 1.2059 | 1938 | 1.4374 | 1972 | 1.6822 |
| 1822 | 1.0157 | 1900 | 1.3024 | 1953 | 1.5953 | 1982 | 1.7163 |
| 1851 | 1.1544 | 1925 | 1.3836 | 1966 | 1.6450 | 1992 | 1.7211 |

**Source**: Shen, 2002.

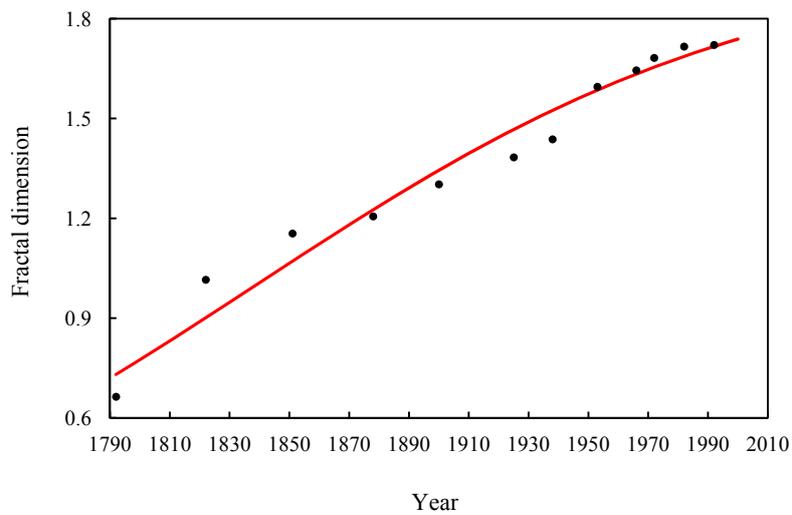

Figure 3 The fractal dimension growth of Baltimore modeled by the logistic function, 1792-1992

## 3.4 Model alternative--Boltzmann's equation

The logistic model can be employed to predict the future or the unknown fractal dimension values and estimate some missing values within the study period. For instance, we can estimate the fractal dimension after 1820 for London, or the fractal dimension after 1935 (especially the missing value in 1962) for Tel Aviv, or the dimension value after 1992 for Baltimore. However, if we use the model to analyze the history of urban evolution, the results seem not to accord with the actual situations. For example, if we use the logistic model to estimate the peak values of fractal dimension growth rate of Tel Aviv in history, the years when the top speed appeared are about 1868 to 1869 for region 1 (the peak value of fractal dimension growth is about 0.0090 per year), 1891 to 1892 for region 2 (the peak value is about 0.0106 per year), and 1897 to 1898 for region 3 (the peak value is about 0.0114 per year), respectively. It is hard to explain these calculations on Tel Aviv. Therefore, we need another model for historical analysis of city development.

In theory, the box dimension varies from 0 to 2, thus the logistic function is a good selection for the future dimension prediction due to its simplicity. However, in practice, the box dimension always comes between 1 and 2 rather than 0 and 2. In this instance, the Boltzmann equation is a good choice for estimating and predicting fractal dimension. Boltzmann's equation for fractal dimension growth can be expressed in the form (Chen, 2012)

$$D(t) = D_{min} + \frac{D_{max} - D_{min}}{1 + (\frac{D_{max} - D_0}{D_0 - D_{min}})e^{-kt}} = D_{min} + \frac{D_{max} - D_{min}}{1 + \exp(-\frac{t - t_0}{p})}, \quad (7)$$

where $D_{min}$ refers to the lower limit of fractal dimension, $D_{max}$ to upper limit of fractal dimension, i.e. the capacity of fractal dimension, $p$ is a scaling parameter associated with the initial growth rate $k$, and $t_0$, a temporal translational parameter indicative of a critical time, when the rate of fractal dimension growth indicating city growth reaches its peak. The scale and scaling parameters can be defined respectively by

$$p = \frac{1}{k}, \quad t_0 = \ln(\frac{D_{max} - D_0}{D_0 - D_{min}})^p.$$

For the normalized data, equation (7) can be rewritten as

$$D^*(t) = \frac{D(t) - D_{min}}{D_{max} - D_{min}} = \frac{1}{1 + (1/D_0^* - 1)e^{-kt}}, \qquad (8)$$

where $D_0^* = (D_0 - D_{min})/(D_{max} - D_{min})$ denotes the normalized result of $D_0$, the initial value of fractal dimension. For equation (1), $D_{min}=0$, thus the above logistic model can also be "normalized" and re-expressed as

$$D^*(t) = \frac{D(t)}{D_{max}} = \frac{1}{1 + (\frac{1}{D_0/D_{max}} - 1)e^{-kt}} = \frac{1}{1 + (1/D_0^* - 1)e^{-kt}}. \qquad (9)$$

This suggests that, for the normalized variable of fractal dimension, the Boltzmann equation is identical in form to the logistic function.

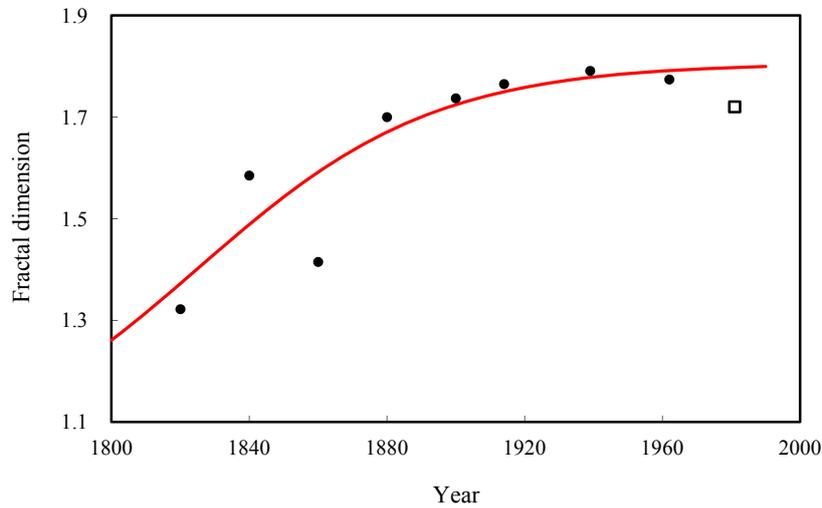

**Figure 4 The fractal dimension change and the trend line predicted by the Boltzmann equation of London, 1760-1990**

The examples of London and Tel Aviv can be employed to make a further analysis by using equation (7). For simplicity, let $D_{min}=1$ (this is not an exact selection), then a least square computation of the data in Table 1 gives a model for London such as

$$D(t) = 1 + \frac{0.806}{1 + \exp(-\frac{t - 5.302}{34.266})}.$$

The goodness of fit is about $R^2=0.847$ (Figure 4). This suggests that the original growth rate of the fractal dimension of London's spatial form is about $k=1/34.266$, and after about $t_0=5.302$ years from the initial year, $n_0=1820$, the fractal dimension value arrived at $D(t)=(D_{max}+1)/2\approx1.403$. In other words, the fractal dimension growth reached its top speed during 1825 and 1826 (1820+5.302). In contrary to this, based on equation (1), or equation (7) with $D_{min}=0$, the years when the peak value of fractal dimension growth rate came between 1773 and 1774 (Figure 5). This seems to be more inconsistent with the real case.

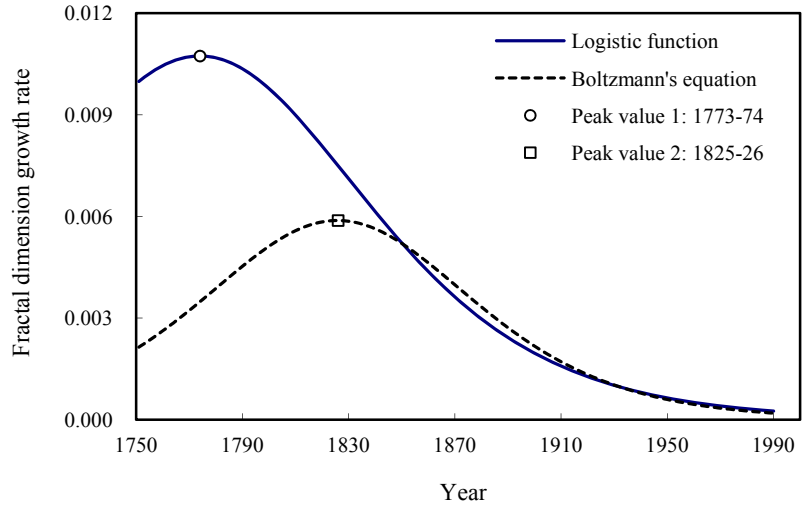

**Figure 5 Two curves of fractal dimension growth rates of London's urban form based on the logistic function and Boltzmann's equation, respectively, 1750-1990**

All the peak values of the *relative growth rates* of population for Inner London (0.0320), Outer London (0.0583), and Greater London (0.0353) appeared during 1851 and 1861 (the time interval of sampling is 10 year). Where the *absolute growth rates* are concerned, the years when appeared the top speed is about 1861 to 1871 for Inter London, 1941 to 1951 for Outer London, and 1901 to 1911 for Greater London (Figure 6). In 1939, Greater London reached its population's peak (8,615,245) (see: http://www.londononline.co.uk/factfile/historical/). Letting $D_{min}=1.25$, we will have $t_0\approx32.929$, and the year of the maximum growth rate of fractal dimension is about 1820+32.929=1852.929. This implies the peak value of fractal dimension growth appeared during 1852 and 1853, corresponding to the relative growth rate of London's population in the mass. A question is why to take $D_{min}=1.25$. In another words, how to objectively determine the lower limit of fractal dimension for a city's form?

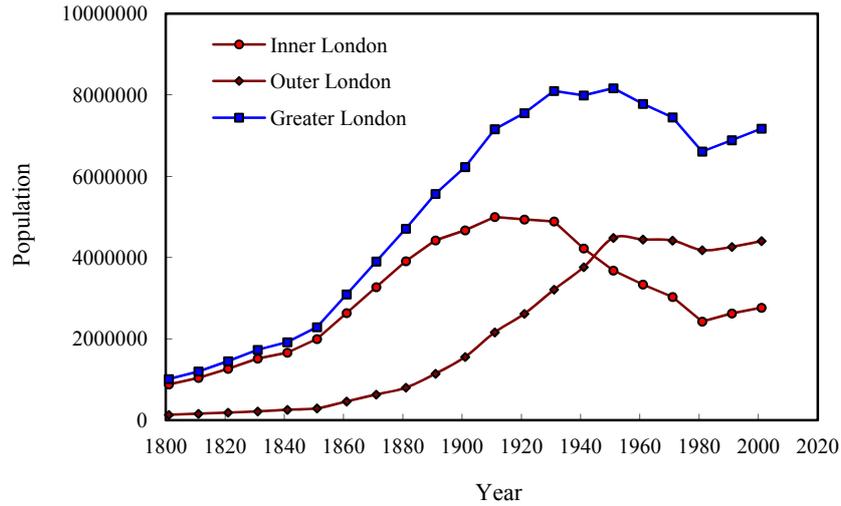

**Figure 6 The population growth curves of London, 1801-2001** (Source: A vision of Britain through time. *Great Britain Historical GIS*. Available from: http://en.wikipedia.org/wiki/Demography_of_London)

The fractal dimension of Tel Aviv can also be modeled with Boltzmann's equation. This seems to correspond to the work of Benguigui *et al* (2001b), who fitted Boltzmann's equation to the population data of Tel Aviv and its satellite towns. Now, letting $D_{min}$=1 and then applying equation (7) to Tel Aviv gives the results of parameter estimation displayed in Table 5. From the scaling parameter $p$ values, we know that the fractal growth rate in region 1 is less than that in region 2, which in turn less than that in region 3 (Figure 7). In short, urban fractal dimension growth in suburbs and exurbs seems to be faster than that in the central part of Tel Aviv. From the scale parameter $t_0$ values, we can estimate that the peak values of fractal dimension growth appeared around 1927 (1935-8. 099), 1940 (1935+4.148), and 1944 (1935+8.723), respectively.

Table 5. The estimated values of the parameters in the Boltzmann equation for Tel Aviv

| Parameter | Parameter value | | |
|---|---|---|---|
| | region1 | region2 | region3 |
| Maximum dimension $D_{max}$ | 2.000 | 1.845 | 1.756 |
| Minimum dimension $D_{min}$ | 1 | 1 | 1 |
| Scaling parameter $p$ | 45.053 | 27.993 | 24.406 |
| Scale parameter $t_0$ | -8.099 | 4.148 | 8.723 |
| Goodness of fit $R^2$ | 0.966 | 0.990 | 0.986 |
| Years of top speed for dimension | 1926-27 | 1939-40 | 1943-44 |
| Peak value of dimension growth rate | 0.0055 | 0.0075 | 0.0077 |

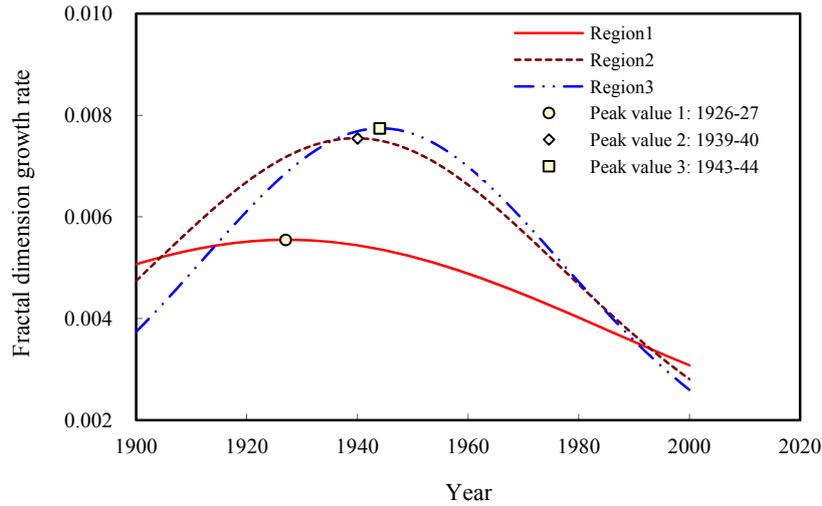

**Figure 7 A spatio-temporal diffusion pattern of Tel Aviv from 1750 to 2000** (Three curves of fractal dimension growth rates are based on Boltzmann's equation of Tel Aviv's urban form in three study regions)

Tel Aviv was the national capital at the time of creation of the State of Israel in 1948. Following the state creation was waves of immigration into the metropolis of Tel Aviv-Yafa (Benguigui *et al*, 2000). In fact, the relative growth rate of Tel Aviv's population seemed to reach its peak before 1948, even before 1940 (see Benguigui *et al*, 2001b for the related data and figures). It is interesting to compare Figure 7 with the geography and history of Tel Aviv. At about 1926 and 1927, the growth rate of fractal dimension of region 1 (central part) reached its peak value (0.0055); at about 1939 and 1940, the fractal dimension growth rate of region 2 (central part and northeast part) got on top (0.0075); and at about 1943 and 1944, the fractal dimension growth rate of region 3 (entire metropolis) culminated (0.0077). Previous studies have suggested that Tel Aviv is a dense city with heterogeneous land use (Frenkel, 2004), and there are differences in the behavior between the central part and northern and southern tiers of this metropolis (Benguigui *et al*, 2000; Krakover, 1985; Shachar, 1975). The sequence of fractal dimension growth seems to be as follows: from the central part to northeast part (form 1926 to 1940) and then to southern part (from 1938 to 1944), from local area to entire metropolis, and as a whole, the city seemed to evolve from a quasi-fractal into a city fractal. The metropolis of Tel Aviv is situated on the Israeli Mediterranean coast, and the city developed along the coastline. Therefore, the initial dimension or the lower limit of fractal dimension is close to $D_{min}=1$. This differs from London, which,

approximately, seems to be of isotropy in spatial diffusion of population (see Batty, 2008 or Batty and Longley, 1994 for the related figures and data of London). As a result, the initial fractal dimension of London's form might be greater than $D_{min}=1$. As for Baltimore, the initial value of fractal is less than 1 owing to the fixed study area, thus the lower limit of fractal dimension can be set as $D_{min}=0$. In this instance, Boltzmann's equation has no difference from the logistic function.

The lower and upper limits of the fractal dimension of urban form are an interesting but a complicated problem, which can be discussed in a specail article. In theory, if a fractal city is defined in a 2-dimension space, then the Lebesgue measure of a fractal set ($F$) of theoretical urban form is $\lambda(F)=0$, and the Euclidean dimension of the embedding space is $d=2$. Therefore, the theoretical value of fractal dimension will come between 0 and 2. However, a real city is not a real fractal, but a pre-fractal (Chen, 2011). The Lebesgue measure of actual urban form is not 0, i.e., $\lambda(F)\neq0$. As a result, the empirical value of fractal dimension will vary from 1 to 2. On the other hand, the empirical fractal dimension value depends on the method of fractal dimension measurement (Batty and Longley, 1994; Chen, 2012; Frankhauser, 1998). If we employ the box-counting method based on the fixed largest box independent of urban shape, the fractal dimension value will vary from 0 to 2; however, if we adopt the box-counting method based on the variable largest box depending on urban shape, the fractal dimension value will range from 1 and 2. As for the upper limit, the theoretical value of $D_{max}$ is less than 2, but the empirical value of $D_{max}$ may be greater than 2. If an empirical box dimension value exceeds the Euclidean dimension of the embedding space of a fractal, the estimated value should be replaced by the Euclidean dimension in order to avoid the logical contradiction.

## 4. Questions and discussion

### 4.1 A generalized logistic model

The difference of social systems from the classical physical systems rests with symmetry breaking of mathematical laws. The laws of the classical physical systems are of translational symmetry in both space and time, while the regularity of social system is not of spatio-temporal translational symmetry (Chen, 2008). Therefore, the mathematical relations and the related parameters are not as stable as those in classical physics. In fact, for many Chinese cities, the common logistic model or Boltzmann'e equation is not applicable. If we fit the logistic function to the sample path from the time series of fractal dimension of urban form, we always cannot find a proper value for the capacity parameter. In other words, the $D_{max}$ value cannot converge at a certain value that is equal to or less than 2, the Euclidean dimension of embedding space. A great number of mathematical experiments show that the fractal dimension growth of urban form of Chinese cities follows a quadratic logistic function:

$$D(t) = \frac{D_{max}}{1+(D_{max}/D_0 - 1)e^{-(kt)^2}} \quad \text{or} \quad D(n) = \frac{D_{max}}{1 + A\exp(-(k(n-n_0)^2))}, \quad (10)$$

in which the notation is the same as in equation (1), and $D_{max} \leq 2$. For the example, for the metropolitan area of Beijing, the capital of China, the box dimension of urban form from 1984 to 2009 can be modeled by equation (10) (Table 6). For the box dimension, the mathematical model is as follows

$$\hat{D}(t) = \frac{1.8627}{1 + 0.3070e^{-(0.0572t)^2}},$$

where $D(t)$-hat refers to the fractal dimension of time $t$. The squared correlation coefficient is about $R^2=0.9955$ (Figure 8). The fractal dimension growth of the metropolitan area corresponds to Beijing's population growth. Fitting the quadratic logistic function to the six times of census data of Beijing's urban population yields a model as below:

$$\hat{P}(t) = \frac{23206725}{1 + 7.3611e^{-(0.0295t)^2}},$$

where $P(t)$-hat denotes the population size of time $t$. The coefficient of determination is about $R^2 = 0.9459$. This suggests that there is inherent relation between urban population and fractal dimension indicating space-filling.

Table 6 The urban area, fractal dimension, and the related measurements of Beijing city (1984-2009)

| Year | Time square ($t^2$) | Urbanized area (A) | Fractal dimension (D) | Filled-unfilled ratio (FUR) | Level of space filling (LSF) |
|---|---|---|---|---|---|
| 1984 | 0 | 365058785.8 | 1.4280 | 0.9846 | 0.4961 |
| 1988 | 16 | 508757623.1 | 1.4465 | 1.0729 | 0.5176 |

| 1991 | 49  | 601060517.8  | 1.4931 | 1.3342  | 0.5716 |
| 1994 | 100 | 752786609.9  | 1.5074 | 1.4279  | 0.5881 |
| 1998 | 196 | 980603624.8  | 1.5967 | 2.2424  | 0.6916 |
| 2001 | 289 | 1375795387.0 | 1.6582 | 3.2170  | 0.7629 |
| 2006 | 484 | 1990590617.0 | 1.7607 | 7.4584  | 0.8818 |
| 2009 | 625 | 2347738581.0 | 1.7888 | 10.6706 | 0.9143 |

**Note**: The measurement filled-unfilled ratio (FUR) and level of space filling (LSF) are defined by Chen (2012).

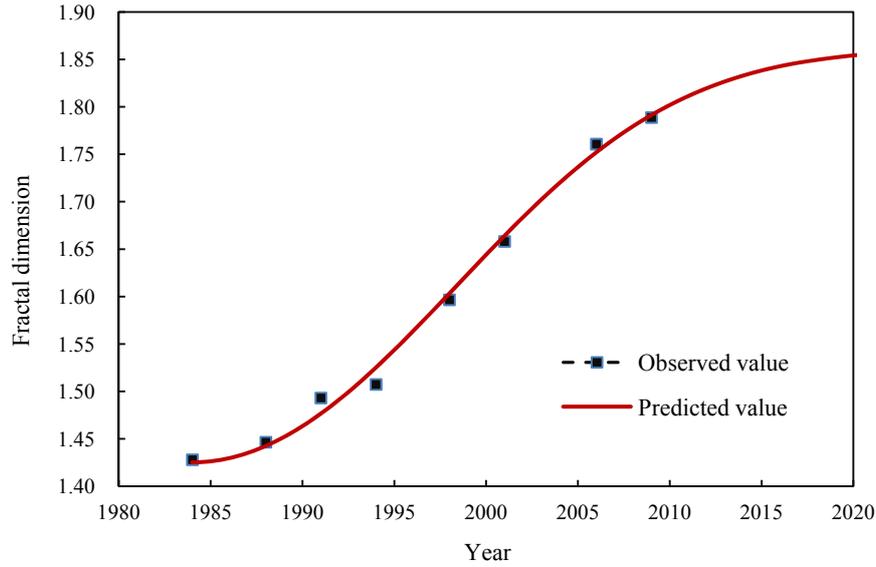

**Figure 8 A curve of fractal dimension growth of Beijing's urban form (1984-2020)**

Now, the logistical model and the quadratic logistic model can be theoretically unified into a fractional logistic function (FLF) in the general form

$$D(t) = \frac{D_{max}}{1+(D_{max}/D_0 - 1)e^{-(kt)^b}} \quad \text{or} \quad D(n) = \frac{D_{max}}{1+A\exp(-(k(n-n_0)^b))}, \quad (11)$$

where $b$ is a temporal scaling exponent of logistic processes and the remaining notation is the same as in equation (1) and equation (10). Obviously, if $b=1$, equation (11) will reduce to equation (1); and if $b=2$, equation (11) will change to equation (10). The parameter $b$ value ranges from 0 to 2, namely, $0 \leq b \leq 2$. In practice, if we cannot find a value coming between 0 and 2 for the capacity parameter ($D_{max}$) of the conventional logistic model, it is often a good choice to adopt the quadratic logistic function. Fractal dimension indicates space-filling extent of urban land use. The urbanized area growth of Beijing can be modeled with a fractional logistic function. Based on the dataset consisting of 8 data points of Beijing's area from 1984 to 2009 (Table 7), a fractional logistic model of urban area growth can be made as follows

$$\hat{A}(t) = \frac{3881716000}{1 + 9.0686e^{-(0.0829t)^{4/3}}},$$

where $A(t)$-hat refers to the urbanized area of time $t$. The squared correlation coefficient is about $R^2=0.9964$. This suggests a possible connection of urbanized area to fractal parameters.

In fact, we can obtain many cases from literature to support the logistic model and quadratic logistic model of fractal dimension growth. A number of new examples stemmed from a developed territory of South America. Sun and Southworth (2013) studied the spatial spread of the developed area comprising human settlements and road networks in Amazon tri-national frontier regions using remote sensing data and fractal geometry. The box-counting method was employed to estimate the fractal dimension of the developed areas involving three administrative regions: the Peruvian state of Madre de Dios (include four sampling subregions), Acre in Brazil (include three sampling subregions), and the department of Pando in Bolivia (include three sampling subregions), from 1986 through 2010 (six years). The results composed 10 sample paths of fractal dimension, in which 9 datasets can be fitted to the logistic function ($b$=1), one dataset can be fitted to the quadratic logistic function ($b$=2) (Chen, 2014). Subregion 1 of Madre de Dios in Peru differs from other 9 subregions. If the fractal dimension growth of a developed area (global region) can be modeled with the logistic function, the fractal dimension change of different parts (local regions) of the developed area can be described with the logistic function, too. For instance, if we examine the subregions of a subregion, say, the northeast part, northwest part, southeast part, and southwest part of subregion 1 in Acre of Brazil, the fractal dimension change also follow the sigmoid curves. By means of the general logistic models, we can draw the outlines of the trajectories of fractal dimension growth and predict the trends of space filling in these regions (Figure 9). The work of Sun and Southworth (2013) contributed good material to logistic modeling of spatial replacement dynamics about the conversion processes from rural to urban and from natural to human systems.

**4.2 The use of the fractal dimension function**

Because of the development of non-Euclidean geometry, the focus of scientific research has shifted from truth finding to model building. I agree with the well-known outstanding scientist, Neumann (1961, page 492), who once said: "The sciences do not try to explain, they hardly even try to interpret, they mainly make models." The basic function of a mathematical model lies in explanation and prediction. As Fotheringham and O'Kelly (1989, page 2) once pointed: "All mathematical modelling can have two major, sometimes contradictory, aims: explanation and prediction." The logistic models of fractal dimension growth, including equations (1) and (10), can be used to explain the process of urban evolution and predict the results of city development where city fractals are concerned. The spatial replacement dynamics based on logistic equation can be employed to explain urban evolution. Moreover, the models can be used to estimate the following parameters or things: what is the maximum fractal dimension, namely, the capacity, $D_{max}$, what is the fractal dimension value of a city in given time (say, 1945 for London), when the largest rate of fractal dimension growth appeared or will appear, and so on. For example, for the year 1962, we know the fractal dimension of region 1 of Tel Aviv, but we know little dimension information of regions 2 and 3. Using equation (1), we can estimate the fractal dimension for the two regions and the results are about 1.587 and 1.517, respectively; using equation (7), we can also estimate the fractal dimension for the two regions such as 1.586 and 1.513 or so.

The logistic models can be used as aid for analyzing when and where a city fractal is. Benguigui et al (2000) once gives a criterion based on standard error for judging fractal city. By the knowledge from statistics, we can derive the relation between the standard error and correlation coefficient in the form

$$\delta = D\sqrt{\frac{1/R^2 - 1}{n-2}}. \tag{12}$$

This suggests that, for given fractal dimension $D$ value, the standard error $\delta$ is equivalent to the goodness of fit, $R^2$ values. However, the $R^2$ value is a statistic for description, not for inference (Batty and Longley, 1994; Shelberg et al, 1982). So, the standard error as well as the goodness of fit seems not to be a good criterion for where and when a city fractal is. In this case, the logistic parameters can be used as an assistant criterion: if the $D_{max}>2$, the study area is too small, and we should define a larger study region in order to find a proper city fractal spectrum over time. On the other, if we can determine the appropriate lower limit value of fractal dimension, $D_{min}$, we can utilize equation (7) to make a space-time diffusion analysis for a city. By defining different scales of study regions, as Benguigui et al (2000) and Feng and Chen (2000) did, we can bring to light how a city develops from the core to the periphery, and when and where a city fractal appears.

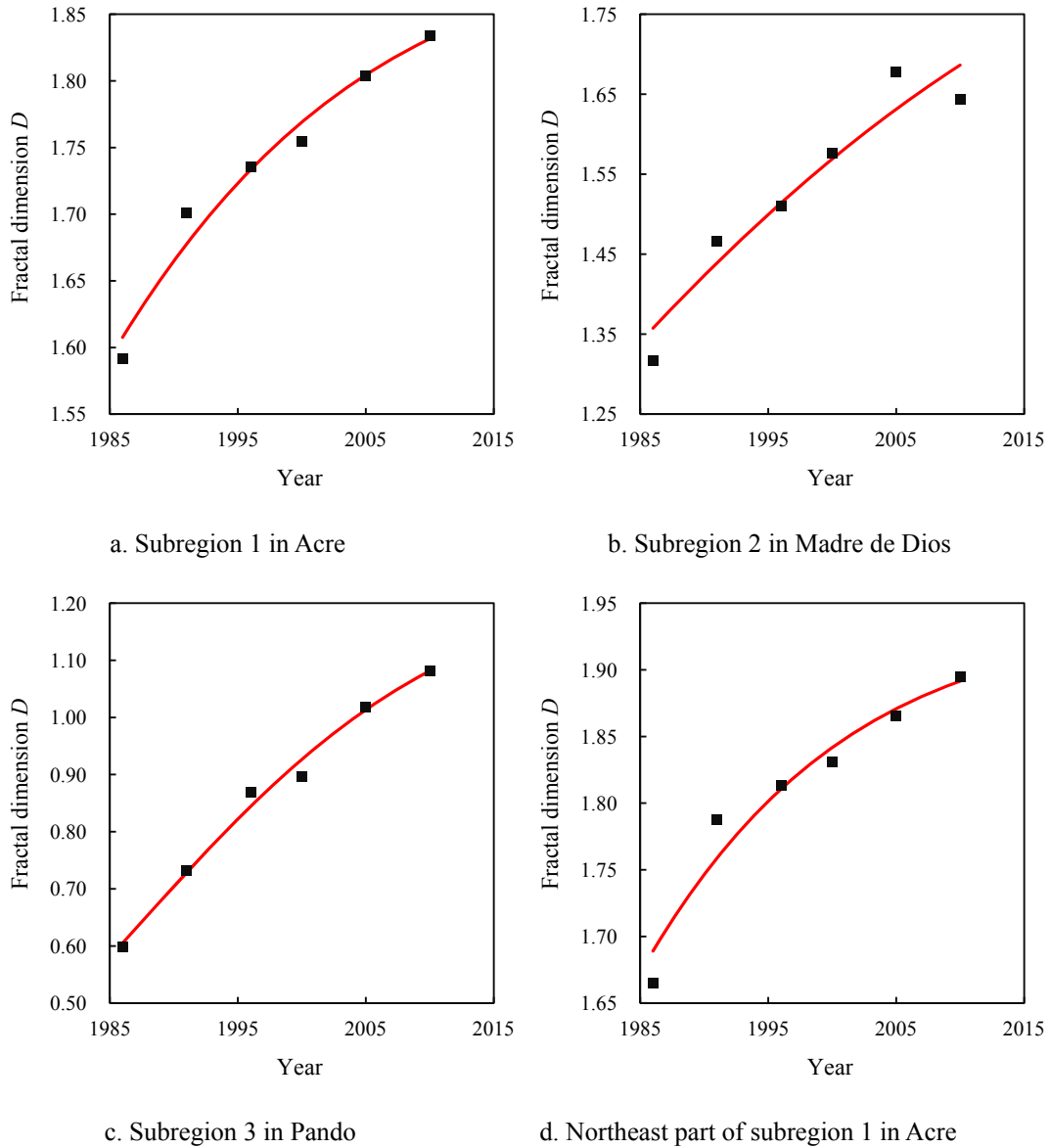

a. Subregion 1 in Acre

b. Subregion 2 in Madre de Dios

c. Subregion 3 in Pando

d. Northeast part of subregion 1 in Acre

**Figure 9 The fractal dimension change and the trend line predicted by partial logistic models of developed areas in Amazon tri‑national frontier regions, 1986-2010**

The logistic models of the fractal dimension growth of urban form can be used to polarize thinking and to pose sharp questions, just like other fractal models of cities (Batty, 1991). As Kac (1969) once observed: "Models are, for the most part, caricatures of reality, but if they are good, then, like good caricatures, they portray, though perhaps in distorted manner, some of the features of the real world. The main role of models is not so much to explain or predict--although ultimately these are the main functions of science--as to polarize thinking and to pose sharp questions." Anyway, the purpose of modeling is insight rather than numbers (Hamming, 1962).

Urban form is associated with urbanization (Knox and Marston, 2009). The intra-urban growth process reflects the process of interurban population flow as well as urban-rural emigration and immigration. Therefore, so far as the mathematical model is concerned, urban growth in intra-urban geography can be compared to the urbanization in interurban geography. For the process of urbanization in the Western countries, the percentage urban population can be described by the logistic function, equation (1). In contrast, the level of urbanization in China cannot be characterized with equation (1). A conjecture is that there is relationship between the fractal dimension growth and urbanization. If so, the time series of level of urbanization of China will follow the quadratic logistic growth. All these studies can be integrated into the general framework of spatial replacement dynamics, which will possibly form one basic theory of theoretical geography (Chen, 2012; Chen, 2014).

## 5. Conclusions

The fractal dimension growth of urban morphology can be described with logistic function according to the squashing principle. The logistic function is in fact a special form of Boltzmann's function. For the normalized data of dimension, Boltzmann's equation can be transformed into a logistic function. The logistic function is simpler and more familiar to geographers than Boltzmann's equation, so this paper is based on the logistic models rather than Boltzmann's model. The logistic model can be used to estimate and predict the unknown fractal dimension. Compared with logistic function, Boltzmann's equation is smarter and can be employed to analyze the history of urban spatial development. However, both the conventional logistic function and Boltzmann's equation are suitable for the fractal cities of the developed countries. As for the cities of developing countries such as China, the fractal dimension growth can be modeled with the quadratic logistic function. The logistic function and the quadratic logistic function can be unified into a fractional logistic model, which is more universal than the traditional form of the logistic function and associating the conventional logistic function with the quadratic logistic function.

For spatial analysis of urban form and growth, the uses and significance of the logistic models of fractal dimension are as follows. First, the models can be used to estimate the past missing fractal dimension values and predict the future unknown fractal dimension, including the capacity parameter ($D_{max}$). Second, the models can be employed to estimate the peak value of the fractal dimension growth rate (d$D$/d$t$). The growth rate of fractal dimension suggests the rate of urban growth in given time. Third, the models can be applied to analyzing where a city fractal is. If the study area is too small, the capacity parameter will go beyond the limit of Euclidean dimension of embedding space ($D_{max}>d$=2). Fourth, the models make a possible new approach to researching urbanization. Urban form is correlated to urbanization, and both the time series of fractal dimension and that of the level of urbanization can be modeled with the generalized logistic function. Fifth, by means of the models, we can research the spatial replacement dynamics. A process of urban growth is actually a series of behaviors of spatial substitution: the rural space is gradually replaced by urban space. The model can be applied to multifractals dimension spectrum of urban form. That is, if the capacity dimension $D_0$ of a city's form follows the logistic or quadratic logistic growth, the other dimension such as information dimension $D_1$, correlation dimension $D_2$ as well as the local fractal dimension $f(\alpha)$ will also follow the logistic or quadratic logistic growth. Owing to the limitation of space, multifractal dimension growth models will be discussed in a companion paper.

## Acknowledgements

This research was sponsored by the National Natural Science Foundation of China (Grant No. 41590843 & 41171129). The support is gratefully acknowledged. I would like to thank Dr. Lucien Benguigui, who gave me a number of constructive comments and suggestions for this paper, and one of my students, Ms Huang, for data processing of Beijing city.